\documentclass[prl,final,preprint,showpacs,floatfix]{revtex4}
\usepackage{dcolumn,longtable}
\usepackage{graphics}
\usepackage{graphicx}
\usepackage{amsmath}
\usepackage{amssymb}
\usepackage{epsfig}

\begin{document}

\title{Designing Conducting Polymers Using Bioinspired Ant Algorithms}

\author{B. V. C. Martins}
\author{G. Brunetto}
\author{F. Sato}
\author{V. R. Coluci}
\author{D. S. Galv\~ao$^*$}
\email{galvao@ifi.unicamp.br}

\affiliation{Instituto de F\'{\i}sica ``Gleb Wataghin'',
Universidade Estadual de Campinas, Unicamp 13083-970, Campinas,
S\~ao Paulo, Brazil}

\date{\today}

\begin{abstract}

Ant algorithms are inspired in real ants and the main idea is to create virtual ants that travel into the space of possible solution depositing virtual pheromone proportional to how good a specific solution is. This creates a autocatalytic (positive feedback) process that can be used to generate automatic solutions to very difficult problems. In the present work we show that these algorithms can be used coupled to tight-binding hamiltonians to design conducting polymers with pre-specified properties. The methodology is completely general and can be used for a large number of optimization problems in materials science.

\end{abstract}

\maketitle

\section{Introduction}

The design of materials with pre-specified properties represents a complex and very difficult problem. Recently, many groups have attempt to use artificial intelligence or automatic methods to design new materials \cite{giro1,giro2, giro3,design,zunger}. 

Real ants consist in social insects that, while individually they can not be considered very intelligent creatures, but as a colony they display remarkable capability of solving very complex optimization problems. This kind of collective intelligence of the swarm \cite{bonabeau1,bonabeau2} can be considered an emmergent property of the system since it appears to be due to the high level of interaction among the agents (ants). The problem of finding food in complex environments is similar to some hard computational problems that can not be handled by conventional searching methodologies. Examples of these problems are present in distinct areas such as protein folding, logistics and the design of materials with predetermined properties.

Based on the capabilities of real ants to solve hard optimization problems (such as finding best or quasi-best paths connecting a food source and the ant nest) a new class of metaheuristic algorithms, named ant system (ANT), was recently developed \cite{dorigo1,dorigo2,bonabeau1,bonabeau2}. The mechanism that allows the ants to establish optimized paths is based on the fact that when ants move they deposit chemical markers (pheromones), which defines specific trails to be travelled. The ants tend to move along trails where the pheromone concentration is higher. This kind of process is known as an autocatalytic one due to the positivie feedback. For instance, if two ants establish two different routes connecting food sources and the nest, the process of going back and forth  along these different routes will increase the amount of pheromone in the most best one (Fig. 1a-1c), since it will be more walked in the same amount of time. Higher concentration of pheromone in one specific trail increases the probability of another ant moving along this trail and so on, generating the autocatalytic process. This is equivalent to a computational parallel processing probing the space of good solutions in a fast and efficient way.

\begin{figure}[ht!]
\begin{center}
\includegraphics[width=5cm]{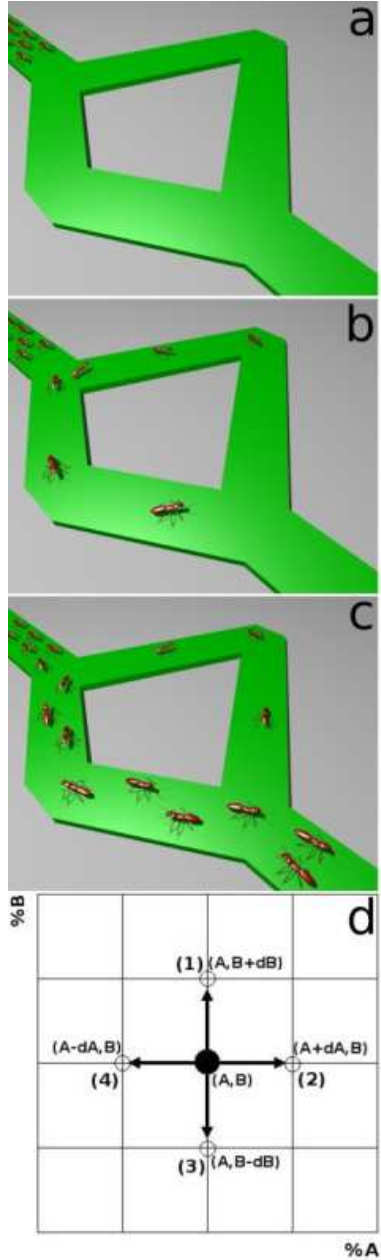}
\end{center}
\caption{Schematic view of the ant algorithms. See text for discussions.}
\label{fig:monomers}
\end{figure}

These aspects of the ant organization were computationally translated into the ant algorithm. The idea is to create artificial (virtual) ants that 'travel' in trails of the phase space of possible solutions. These ants can 'communicate' through the virtual 'pheromones' deposited on these trails, with the deposition rate proportional to the effectiveness of the tentative solution (Fig. 1b). As in the case of real ants the positive feedback in the virtual space of solutions is capable of generating good solutions that 'autocatalitically' evolve in time, in general being able to solve very hard optimization problems \cite{dorigo1,dorigo2,bonabeau1,bonabeau2}. The ant algorithm implicitly possessing abilities of 'memory' and 'learning' can outperform even recent bioinspired class of genetic algorithms \cite{holland}.
   
Genetic algorithms consist in a class of bioinspired algorithms that were originated from the studies of J. Holland in the 1970s \cite{holland}. The metaphor underlying this method is that of natural evolution, whose concepts are followed in a very simple scheme that allows computers to evolve automatic solutions over time. The method is constructed based on the evolution of populations using processes analogues to the biological ones, namely the mutation and the crossover (variation, selection, and inheritance). Genetic algorithms are very effective to obtain good solutions from ``scratch'', but they are less efficient to improve on good ones. The reasons are the same as in real biology. As the available genetic space starts to exhibit little variations, the mechanisms of evolution (in special selection) are no longer very effective and new solutions must heavily depend on the mutations, which in general are effective only over long periods of time. Although there are some computational tricks to overcome these difficulties and to improve on the algorithm performance, sometimes they are very difficult to implement numerically. Thus, it would be desirable to have alternative methodologies. The class of ant algorithms can be one of these alternatives.

In the present work we adapted ant algorithms to solve the problem of designing conducting polymers. Polymers constitute a very large and important class of materials that exhibit a rich variety of mechanical and electronic properties, varying from soft to very hard, from insulators to high conducting materials \cite{polymers}. Due to the richness of the carbon atom reactivity an almost infinite number of new structures is possible. This makes the systematic search for new structures impossible and the trial and error approach has been the rule. In this sense the use of combinatorial and/or artificial intelligence methods to assist in the design of new or to improve on the existing materials would be very usefull. In order to show that ant based algorithms can be an effective tool to do this we illustrate its application to the complex problem of designing polymeric alloys with pre-defined electronic properties. We considered the case of the polyaniline (PANi) family, one of the most studied conducting polymers \cite{pani}. PANi encompass a family of compounds where nitrogen atoms connect six-membered carbon rings of benzenoid or quinoid character. They can have three different monomeric units (PAN1,PAN2,PAN3, see Fig.2) depending on its degree of oxidation and/or protonic doping. The relative concentration of these units in the polymeric chain define its conductivity, varying from insulator to highly conducting material \cite{pani,galvao1,galvao2,galvao3}. Considering that, in general, each polymeric chain contains hundreds of PAN1, PAN2, PAN3 units \cite{pani} a  huge number of possible configurations varying the relative concentration of the different units is possible. What are the concentrations of PAN1, PAN2, PAN3 to produce the most conductive alloy (metallic states) ? The metallic state is characterized by a zero gap value and for delocalized states at the Fermi level. We coupled our based ant algorithm to the simple Huckel hamiltonian (one of simplest tight-binding models)\cite{huckel} to solve this problem. We were able to automatically generate very good solutions. The methodology is completely general and can be coupled to sophisticated hamiltonians (even \textit{ab initio} ones) and can be applied to design other classes of materials such as metals or oxides.

\begin{figure}[ht!]
\begin{center}
\includegraphics[width=7cm]{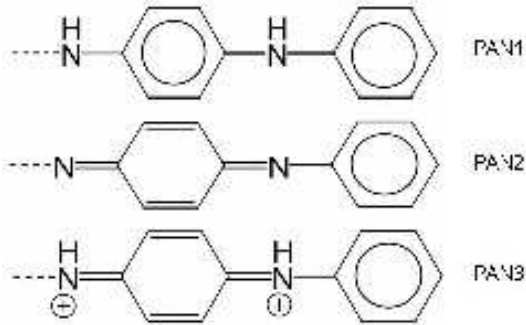}
\end{center}
\caption{Structure of the three monomeric units used in the calculations: PAN1, PAN2, PAN3.}
\label{fig:monomers}
\end{figure}

\section{Methodology}
\subsection{Ant System}
Our adapated ant algorithm is based on two principles: positive and negative feedback, which are associated with the creation and destruction of trails, respectively. The trail creation occurs when an ant walks between two points on a potential surface $P(x,y)$,  depositing an amount of pheromone that is proportional to the quality of the solution at the end point. This process is described by the trail function: 
\begin{equation} \label{eqn:rule}
T(x_1,y_1,x_2,y_2,t) = T(x_1,y_1,x_2,y_2,t-1) + P(x_2,y_2)
\end{equation}
where $(x_1,y_1)$ and $(x_2,y_2)$ are the the starting and end points (see Fig. 1d), respectively, and $t$ is the current time step. The potential function $P(x,y)$ is the mathematical translation between this method and the problem to be solved. This function is the responsible for the positive feedback of good solutions since better points will generate better (more walked) trails. The trail is the most important interaction feature in this method since it is the only communication mechanism among the ants. 

The contraposition to this pheromone accumulation in certain points is given by the regulatory negative feedback contribution, introduced as a constant evaporation rate term, $\rho < 1$, acting over the trail function:
\begin{equation} \label{eqn:evaporation}
T(x_1,y_1,x_2,y_2,t) = \rho T(x_1,y_1,x_2,y_2,t).
\end{equation}

This term is the responsible for the removing of the trails that are not being walked, since in these trails there is no pheromone redeposition. This is a strong selection acting over trails leading to bad solutions, contributing to the restriction of the circulation in the surface and to the saving of time in the searchings. 

This pheromone deposition/evaporation process has the only objective of guiding the ant movement choices when faced with the options given by potential surface. The choices are quantified by an evaluation function:
\begin{equation} \label{eqn:choice}
E(x_1,y_1,x_2,y_2,t) = a P(x_2,y_2) + b T(x_1,y_1,x_2,y_2,t) + c R(t)
\end{equation}
which is calculated for each of the possible neighboring final points connected to the current point occupied by the ant. The tuning coefficients  $a$, $b$ and $c$ are used to calibrate the relative weights between the contributions. $R(t)$ is a random term which represents a kind of noise or uncertainty in the information acquired from the environment, leading to the appearance of small fluctuations which are relevant in situations where two different evaluation functions have very close values. The point with the highest value for the evaluation function is the one which will be the next to be visited. 

\subsection{Polymer Problem}

As mentioned before our objective is the determination of the best set of relative concentrations ($x(PAN1)$,$y(PAN2)$,$z(PAN3)$) (see Fig. 2) that produces the polymer with better electrical conductivity. One of the objectives here is to demonstrate that ant algorithms can solve this kind of problem in a fast and efficient way without extensively probing the space of possible solutions. Thus, we started analyzing a size problem that can be solved by systematic search and then we compare the 'exact' solutions with the ones obtained using the ant algorithms. In this sense we considered PANi chains containing 100 (equal or different) units. In order to obtain their electronic structure we used the simple H\"uckel (tight-binding model) hamiltonian \cite{huckel,galvao1,galvao2,galvao3}:

\begin{equation} \label{eqn:huckel2}
H=\sum_{i=1}^{n} \alpha_{i} \vert i \rangle \langle i \vert + \sum_{i,j,i \neq j} \beta_{i,j} \vert i \rangle \langle j \vert,
\end{equation}
where  $\alpha_{i}$ and $\beta_{i,j}$ represents the Coulomb and the hopping integrals, respectively \cite{huckel}. The wavefunctions are written as a linear combination of $n$ atomic $\pi$-orbitals $\phi_{l}$
\begin{equation} \label{eqn:huckel1}
\vert i \rangle =\sum_{l=1}^{n} c_{il} \vert \phi_{l} \rangle,
\end{equation}

The degree of electronic wavefunction delocalization can be estimated from the inverse participation number (IPN) \cite{ipn} given by:

\begin{equation} \label{eqn:ipn}
\displaystyle \mathrm{IPN}=\frac{\sum_{l=1}^{n}|c_{l}|^{4}}{(\sum_{l=1}^{n}|c_{l}|^{2})^{2}}.
\end{equation}

where $c_{l}$ are the coefficients of the highest occupied molecular orbital(HOMO)(Fermi level).

IPN can assume values from 0 (maximum delocalization, the wavefunction is spread over the whole polymeric chain) up to 1 (maximum localization, the wavefunction is localized over a single atom). From the IPN and gap (energy difference between the frontier (last occupied - first unoccupied) molecular orbitals) we can characterize the metallic states. In order to be a good conductor the gap value should be smaller than the room temperature thermal energy (25 meV) and the IPN value close to zero. We need do determine the x,y, and z values that satisfy the gap and IPN conditions. The problem can be simplified if we use the constraint relation $x + y + z = 100$, which reduces the space of possible solutions from a tridimensional (x, y, z) to a two-dimensional (x,y surface) one. If the polymeric chains are long enough the relative concentration of the different units is more important than their specific position in the chains, considering that they are randomly formed \cite{galvao1,galvao2,galvao3}. In this case we would have a grid of 100x100 of possible configurations that by the symmetry imposed by our contraint relation is then reduced to a grid of 5151 points. Each point ($x_k,y_k$) ($k=1,\ldots,$ 5151) in the grid represents a possible solution (with specific gap and IPN values) of our problem associated with a polymeric chain containing 100 monomeric units distributed accordingly to the relative concentrations ($x_k,y_k$,$z_k$). We then proceed to calculate the gap and IPN values for each one of these 5151 configurations.
The calculated space of solutions can be better visualized through a potential function of the grid points $P(x_k,y_k)$:

\begin{equation} \label{eqn:potential}
P(x_k,y_k) = log \left( \dfrac{1}{\mathrm{GAP}(x_k,y_k)\times\mathrm{IPN}(x_k,y_k)} \right)
\end{equation}

\begin{figure}[ht!] \centering
\includegraphics[width=6cm]{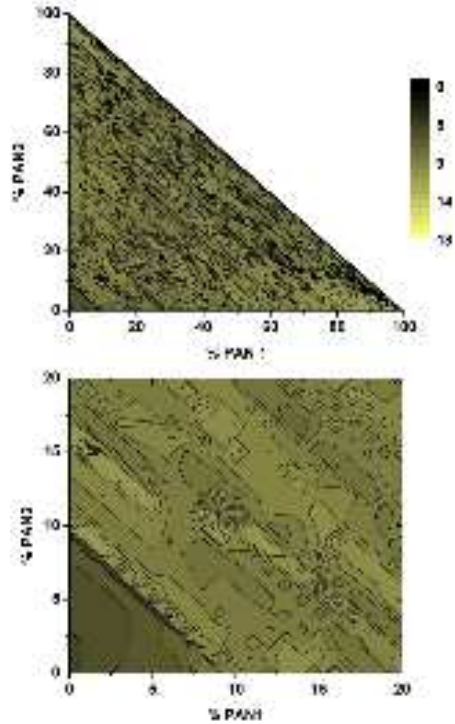}
\caption{Top: Contour plot of the potential grid for PANi chains containing 100 units as a function of the percentage of units x(PAN1)  and y(PAN2); Bottom: Zoomed contour plot for the initial region of the grid. The corrugation of the grid and the existence of many local maxima are indications of the complexity of the potential profile.}
\label{fig:matrix}
\end{figure}

We used a \textit{log} function due to the large variations found for neighboring points. The surface topography of the grid points calculated with this function is presented in Figure \ref{fig:matrix}. The observed triangular shape is a consequence of the used constraint relation. Higher peaks are indication of metallic signatures. The visual complexity of this profile gives an idea of the difficulty of the problem for standard optimization algorithms that in general fail for problems where many local minima and multiple local maxima are present. 

\clearpage

\subsection{Adapted Ant Algorithm}

Our adapted ant algorithm can be coupled to the H\"{u}ckel hamiltonian \cite{huckel} to carry out the automatic search for the metallic structures. The general algorithm fluxogram has the following form: 

\begin{scriptsize}
\begin{eqnarray*}
&&\textrm{To generate the starting positions}\\
&&\mathbf{do} \textrm{ i=1, number of cycles}\\
&&\qquad \mathbf{do} \textrm{ j=1, number of steps}\\
&&\qquad \qquad \mathbf{do} \textrm{ k=1 , number of ants}\\
&&\qquad \qquad \qquad \mathbf{do} \textrm{ l=1,4 (neighbours)}\\
&&\qquad \qquad \qquad \qquad \mathbf{if} \textrm{the next point is permitted}\\
&&\qquad \qquad \qquad \qquad \qquad \mathbf{if} \textrm{ the electronic properties of this point were}\\
&&\qquad \qquad \qquad \qquad \qquad \textrm{not calculated}\\
&&\qquad \qquad \qquad \qquad \qquad \qquad \textrm{Calculate the electronic properties}\\
&&\qquad \qquad \qquad \qquad \textrm{Calculate the potential function}\\
&&\qquad \qquad \qquad \qquad \textrm{Calculate the evaluation function}\\
&&\qquad \qquad \qquad \qquad \mathbf{end\:do}\\
&&\qquad \qquad \qquad \textrm{Choose the next point}\\
&&\qquad \qquad \qquad \textrm{Move the ant}\\
&&\qquad \qquad \qquad \textrm{Update the pheromone trail}\\
&&\qquad \qquad \qquad \textrm{Check if the stop condition was achieved}\\
&&\qquad \qquad \qquad \mathbf{end\:do}\\
&&\qquad \qquad \textrm{Multiply the trail function by the evaporation term}\\
&&\qquad \qquad \mathbf{end\:do}\\
&&\qquad \textrm{To generate the ants positions for the next cycle}\\
&&\mathbf{end\:do}\\
\end{eqnarray*}
\end{scriptsize}

The algorithm is started with the random sorting of the ant starting positions. There are two temporal variables: the cycles and the steps. The cycle is a fixed number of steps after what the ants are sorted again over the grid. Given a point in this grid the ant has four possible directions (up, down, left, right) (see Fig. 1d). In order to avoid closed cycles, each ant stores the sequence of points visited and these are prohibited to be visited again by the same ant. There are other two prohibitions for the ant movements: if there is another
ant in the final point (since two ants are not allowed to have the same coordinates at the same time) and if the final point is outside the region defined by the constraint relation. Everytime when the electronic parameters of a given point are calculated the information is stored in a databank accessible to all ants. This prevents redundant calculations and optimize the searching time. If no improvement on the best solution found is obtained after the maximum number of allowed cycles the simulation is ended. 

The ant system was tested in 100 runs with different initial positions for the ants and different seeds for random number generator. We have used 50 ants, 100 steps, and 5 cycles. The tuning parameters were $a = 1.0$,
$b = 10$, $c = 0.1$, and $\rho = 0.95$  (see Eqs. 2 and 3), and were chosen based on previous testing calculations. The number of ants is approximately 1\% of the total number of grid points, condition that we established as a rule of thumb. The algorithm seems to be very robust. In test simulations the obtained results were not very sensitive to the tuning parameters.

The strategy to quantify the efficiency of the method is not based on the number of steps but on the average number of electronic calculations $N$ over the independent 100 runs necessary for finding a specific grid point, since this is the main time consuming operation in the simulations. As mentioned earlier more then one ant can pass each point, but this does not imply in a repeated electronic calculation, since that for the grid point probed the gap and IPN values were already stored. $N$ in combination with the percentage of grid points probed in the set of simulations ($N_f$), can provide a good measure of the efficiency of the methodology.

\section{Results and Discussions}

In Table \ref{table:points}  we present a summary of the obtained results. The composition (number of PAN1, PAN2, PAN3 units), gap and IPN values for the 10 best solutions obtained from the systematic search are displayed. All these points have effectively a zero gap value since these energies are smaller than the thermal energy at 300 K (25 meV). The associated IPN values are very small, thus indicating a significant electronic delocalization at the Fermi level. Thus, all these 10 points satisfy the required conditions for metallic states. As the gap values are pratically zero and  the IPN difference between the best and the worst case is only $5.8 \times 10^{-4} eV$, from a practical point of view these ten best grid points are basically equivalent.

The results obtained from the ant algorithm: the potential function $P(x_k,y_k,z_k)$, the percentage of the 100 runs where the specific grid point was located ($N_f$) and the average total number of points ($N$) probed are also displayed in Table \ref{table:points}. 

We can see from the Table that the majority of the 10 best points were located in the search runs and in average less thant $30 \%$ of the space of solutions was probed. The best 3 solutions were located in about $80 \%$ of the runs. This is a remarkable performance considering the existence of multiple minima and maxima and that these points are basically equivalents (very close gap and IPN values).

\begin{table}
\caption{Values for GAP, IPN, P, $N_f$, and $N$ for the ten best grid points.}
\begin{center}
\begin{tabular}{ c c c c c c}
\hline
$(x_k,y_k,z_k)$ & GAP($10^{-6}$eV) & IPN($10^{-2}$eV) & $P(x_k,y_k,z_k)$ & $N_f$ (\%) & $N$\\
\hline
(93,5,2) & $0.67$  & $1.072$  & $18.745$ & $88$ & $1768$\\	
(91,6,3) & $0.74$  & $2.009$  & $18.025$ & $79$ & $1557$\\
(95,2,3) & $19.58$ & $0.167$  & $17.236$ & $88$ & $1547$\\
(9,90,1) & $7.24$  & $0.618$  & $16.922$ & $38$ & $999$\\
(14,84,2)& $23.46$ & $0.259$  & $16.622$ & $82$ & $1431$\\
(40,59,1)& $4.97$  & $1.302$  & $16.553$ & $47$ & $1148$\\
(41,58,1)& $4.97$  & $1.302$  & $16.553$ & $58$ & $1265$\\
(38,61,1)& $4.97$  & $1.302$  & $16.552$ & $79$ & $1433$\\
(39,60,1)& $4.97$  & $1.302$  & $16.552$ & $79$ & $1489$\\
(42,57,1)& $4.99$  & $1.130$  & $16.548$ & $63$ & $1350$\\
\hline
\end{tabular}
\label{table:points}
\end{center}
\end{table}

An example of the trail evolution of one simulations is presented in Figure 4. The frame sequence starts with the initial random ant distribution (Figure 4a). In the first steps the pheromone term contribution (see Eqs. 2 and 3) is not yet important since there is pratically no trail overpositions. The situation changes when the ants start to find 'weak' trails formed by single ants. Due to the importante of pheromone contribution to the evalution function (b=0, see Eq. 7), there is a tendency of the of ant to follow these starting trails (Figure 4b). In fact as more ants find these trails, larger will be the amount of pheromone deposited and greater the importance of these paths (Figure 4c). The global evolution of the system goes in a direction where the interconnected trails form a network of preferential paths accessible to all ants, and the positive/negative feedback established (Figure 4d).

\begin{figure}[ht!] \centering
\includegraphics[width=6cm]{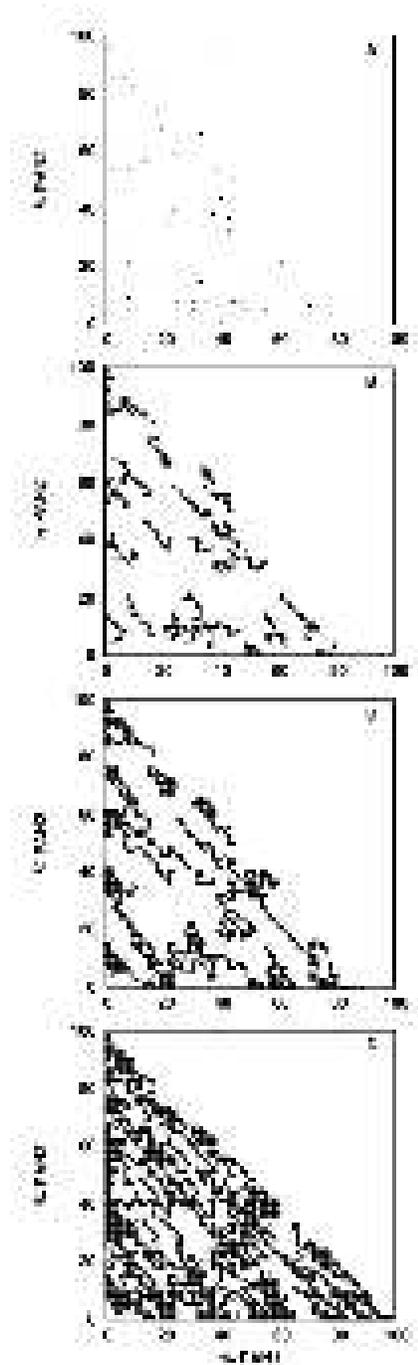}
\caption{Evolution of the trail distribution with time. (A) Initial random sorting; (B) Evolution before the start of trail overlap (20 steps); (C) Formation of the network due to the trail overlap (90 steps); (D) Final set of trails formed (25 steps of cycle 4).}
\label{fig:dynamics}
\end{figure}

\clearpage

\begin{figure}[ht!] \centering
\includegraphics[width=6cm]{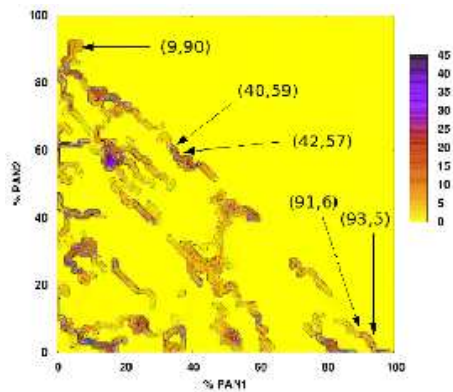}
\caption{Pheromone distribution map at the end of the calculation of Fig. \ref{fig:dynamics}. The five marked points were selected from the top 10 list (Table  \ref{table:points}).}
\label{fig:final}
\end{figure}

However, this is not a static network since the pheromone evaporation continuously acts in the removal of trails. The continued existence of one trail in the long range depends on the quality of the points contained in it. This effect is observed when the final distribution of pheromone is plotted superimposed with some of the 10 best points shown on Table \ref{table:points}. Figure \ref{fig:final} shows this final distribution at the end of the calculation shown in Figure \ref{fig:dynamics}.

It can be noticed that all points showed are in trails that survived after evaporation. This demonstrates that the evaporation (negative feedback) acts mainly in removing bad points where the ant rarely travels. This also shows why the method is so effective: the trails passing through some of the best points are continuously reinforced (positive feedback) and the random term (last term of Eq. 3) warrants the possibility of finding yet better solutions, as the case of real ants. A better idea of the dynamics of these processes can be obtainded from the movie showing an animated sequence of the virtual ant walkings into the space of solutions \cite{video}.

\section{ Summary and Conclusions}

The design of materials with pre-specified properties represents a complex and very difficult problem.  Ant algorithms are a new class of bioinspired algorithms capable of solving very difficult optimization problems. We present here a new methodology for automatic design of polymeric alloys with pre-especified properties, in our case we are looking for metallic signatures. We coupled an adapted ant algoritms to the simple H\"uckel hamiltonian (one of simplest tight-binding methods) in order to determine the composition of polymeric alloys based on polyaniline polymers capable of exhibiting metallic behavior.
Our results show that the ant algorithm is efficient and well suited for the optimization problem involving the design of polymeric alloys. We were able to obtain very good solutions for the PANi alloys, probing less than $30 \%$ of the space of solutions. Preliminary results using the same methodology for ternary and quartenary alloys of very long polymeric chains, where the systematic search is computationally cost prohibitive, are very encouraging. Our methodology is hamiltonian independent, it could be used with other hamiltonians, even \textit{ab initio} ones. It could be applied to other classes of materials, such as metals or oxides. We hope the present work to estimulate further works along these lines.

The authors acknowledge financial support from IMMP/MCT, THEO-NANO and the Brazilian agencies FAPESP, CNPq and CAPES.

\end{document}